\pdfoutput=1
\documentclass[%
reprint,
superscriptaddress,
bibnotes,
amsmath,amssymb,
pra,
floatfix,
]{revtex4-1}
\usepackage{graphicx}
\usepackage{dcolumn}
\usepackage{bm}
\usepackage{hyperref}

\usepackage{natbib}
\usepackage{mathrsfs,mathtools,color,wasysym,dsfont,here}
\usepackage{amsthm}
\usepackage{physics}
\usepackage[all]{xy}
\usepackage{tikz}
\usepackage{amsfonts}
\usepackage{array,booktabs}
\usepackage{comment}
\usetikzlibrary{calc}
\usetikzlibrary{arrows}
\usetikzlibrary{decorations.markings,decorations.pathmorphing}
\usepackage{cases}

\AtBeginDocument{
\heavyrulewidth=.08em
\lightrulewidth=.05em
\cmidrulewidth=.03em
\belowrulesep=.65ex
\belowbottomsep=0pt
\aboverulesep=.4ex
\abovetopsep=0pt
\cmidrulesep=\doublerulesep
\cmidrulekern=.5em
\defaultaddspace=.5em
}

\preprint{}

\begin{document}

\title{Exact eigenstates of extended SU($N$) Hubbard models: Generalization of $\eta$-pairing states with $N$-particle off-diagonal long-range order}

\author{Hironobu Yoshida}
\email{hironobu-yoshida57@g.ecc.u-tokyo.ac.jp}
\affiliation{Department of Physics, Graduate School of Science, The University of Tokyo, 7-3-1 Hongo, Tokyo 113-0033, Japan}

\author{Hosho Katsura}
\affiliation{Department of Physics, Graduate School of Science, The University of Tokyo, 7-3-1 Hongo, Tokyo 113-0033, Japan}
\affiliation{Institute for Physics of Intelligence, The University of Tokyo, 7-3-1 Hongo, Tokyo 113-0033, Japan}
\affiliation{Trans-scale Quantum Science Institute, The University of Tokyo, 7-3-1, Hongo, Tokyo 113-0033, Japan}
\begin{abstract}
We consider $N$-particle generalizations of $\eta$-pairing states in a chain of $N$-component fermions and show that these states are exact (high-energy) eigenstates of an extended SU($N$) Hubbard model. We compute the singlet correlation function of the states and find that its behavior is qualitatively different for even and odd $N$. When $N$ is even, these states exhibit off-diagonal long-range order in $N$-particle reduced density matrix. On the other hand, when $N$ is odd, 
the correlations decay exponentially with distance in the bulk, but end-to-end correlations do not vanish in the thermodynamic limit.
Finally, we prove that these states are the unique ground states of suitably tailored Hamiltonians.
\end{abstract}

\maketitle

\section{Introduction}The off-diagonal long-range order (ODLRO) in the reduced density matrix characterizes a quantum phase in many-particle systems~\cite{yang_concept_1962}. 
For bosons, ODLRO in the one-particle reduced density matrix is a signature of Bose-Einstein condensation~\cite{penrose_cxxxvi_1951,penrose_bose-einstein_1956,dalfovo_theory_1999,leggett_bose-einstein_2001,bloch_many-body_2008,tasaki_spontaneous_2020}, while for fermions, ODLRO in the two-particle reduced density matrix signals superconductivity~\cite{yang_concept_1962,sewell_off-diagonal_1990,nieh_off-diagonal_1995,Tian1992,shen_exact_1993}.

In his seminal paper~\cite{yang__1989}, C. N. Yang found exact eigenstates of the SU(2) Hubbard model called $\eta$-pairing states which possess 
ODLRO in the two-particle reduced density matrix. Although these states are not ground states, there is renewed interest in them in the context of superconductivity and superfluidity~\cite{zhai_two_2005,kitamura_eta_2016,kaneko_photoinduced_2019,buvca_eta_2019,tindall_eta_2019,li_eta-pairing_2020,naoto_tsuji_tachyonic_2021, nakagawa_eta_2021}, and quantum many-body scars~\cite{mark_2020,moudgalya_2020}.

So far, $\eta$-pairing states have been discussed mainly in the SU(2) Hubbard model~\footnote{Recently, extensions of $\eta$-pairing states to models with spin-orbit coupling 
have been discussed. See Refs.\cite{moudgalya_2020,li_eta-pairing_2020}.}. Recently, $N$-component 
Fermionic systems with SU($N$) symmetry have been realized in optical lattices~\cite{Abraham1997,Bartenstein2005,Fukuhara2007,Ottenstein2008,Huckans2009,
Taie2010,Taie2012,Desalvo2010,Lewenstein2012,Scazza2014,Zhang2014,Cazalilla2014,Pagano2014,Hofrichter2016}. Such systems are described by the SU($N$) Hubbard model, which is a generalization of the Hubbard model for the internal degrees of freedom. In a continuous system, Cooper triples, a generalization of Cooper pairs, have been proposed for SU(3) fermions~\cite{tajima_cooper_2020,akagami_condensation_2021}. However, its lattice counterpart is missing. Also, to our knowledge, the generalization of $\eta$-pairing states to $N$-component fermionic systems has yet to be investigated.

$\eta$-pairing states can be represented by a superposition of states in which each site is 
either empty or occupied by two fermions with opposite spins.
Thus, a natural generalization of 
such a state to the case of $N$-component fermions is a superposition of states in which each site is either empty or occupied by $N$ fermions with different flavors.
When the number of fermions is a multiple of $N$, one can construct a series of such states, which we dub {\it $\eta$-clustering states}.
These states are not exact eigenstates of the SU($N$) Hubbard model.
Nevertheless, we can construct a class of modified models in one dimension where these states are zero-energy eigenstates, as shown in this paper. These eigenstates are in the middle of the spectrum, but their entanglement entropy obeys a sub-volume law. Thus, they can be thought of as scar-like states. We also compute the singlet correlation function, which is an $N$-particle generalization of the pair correlation function. When $N$ is even, the states have 
$N$-particle ODLRO because the singlet correlation function 
does not decay to zero at large distances.
On the other hand, when $N$ is odd, the states do not exhibit 
$N$-particle ODLRO because the singlet correlation function decays 
exponentially with distance in the bulk. Interestingly, however,
end-to-end correlations do not vanish in the thermodynamic limit.
Finally, we prove that these states are the unique ground states of suitably tailored Hamiltonians.

\medskip

\section{The model and $\eta$-clustering states}

\subsection{Hamiltonian}
We consider a chain of $N$-component fermions with $L$ lattice sites. For each site $x =1,\ldots, L$, we denote by $\hat{c}^\dagger_{x,\sigma}$ and $\hat{c}_{x,\sigma}$ the creation and annihilation operators, respectively, of a fermion with flavor $\sigma=1, \ldots, N$. We write the normalized vacuum state annihilated by all $\hat{c}_{x,\sigma}$ as $\ket{0}$. The whole Fock space $\mathcal{V}$ is spanned by states of the form $\{ \prod_{x=1}^L \prod_{\sigma=1}^N (\hat{c}^\dagger_{x,\sigma})^{n_{x,\sigma}} \} \ket{0}$ $(n_{x,\sigma}=0,1)$. The number operators are defined as $\hat{n}_{x,\sigma} =\hat{c}^\dagger_{x,\sigma} \hat{c}_{x,\sigma}$ and $\hat{n}_x=\sum_{\sigma=1}^{N} \hat{n}_{x,\sigma}$.
We write $\hat{\eta}^\dagger_x = \hat{c}^\dagger_{x,1}\hat{c}^\dagger_{x,2}\ldots \hat{c}^\dagger_{x,N}$ and $\hat{\overline{c}}^\dagger_{x,\sigma}=[\hat{c}_{x,\sigma},\hat{\eta}_x^{\dagger}]_{\pm}$, where $[~,~]_{\pm}$ denotes the commutator (anticommutator) when $N$ is even (odd)~\footnote{When $N=2$, $\hat{{\overline{c}}}_{x, \sigma}=-\sigma \hat{c}_{x, -\sigma}$, in which case  $\hat{H}_\mathrm{OBC}$ defined as Eq. \eqref{eq:ham} 
reduces to the standard Hubbard model}.

Let us consider the Hamiltonian of the extended SU($N$) Hubbard model with open boundary conditions,
\begin{align}
  \hat{H}_\mathrm{OBC}&=\hat{H}_1 +\hat{H}_{N-1}+\hat{H}_U,  \label{eq:ham} \\
  \hat{H}_1&=-t\sum_{x=1}^{L-1} \sum_{\sigma=1}^{N}(\hat{c}^\dagger_{x,\sigma} \hat{c}_{x+1,\sigma}+\text{h.c.}), \label{eq:onehop}\\
  \hat{H}_{N-1}&=-t \sum_{x=1}^{L-1} \sum_{\sigma=1}^{N}(\hat{\overline{c}}^\dagger_{x,\sigma} \hat{\overline{c}}_{x+1,\sigma}+\text{h.c.}), \label{eq:Nhop}\\
  \hat{H}_U&=U\sum_{x=1}^{L} \hat{n}_x\left(\hat{n}_x-N\right).
  \label{eq:hamint}
\end{align}

\begin{figure}
  \centering
  \includegraphics[width=1.0\linewidth]{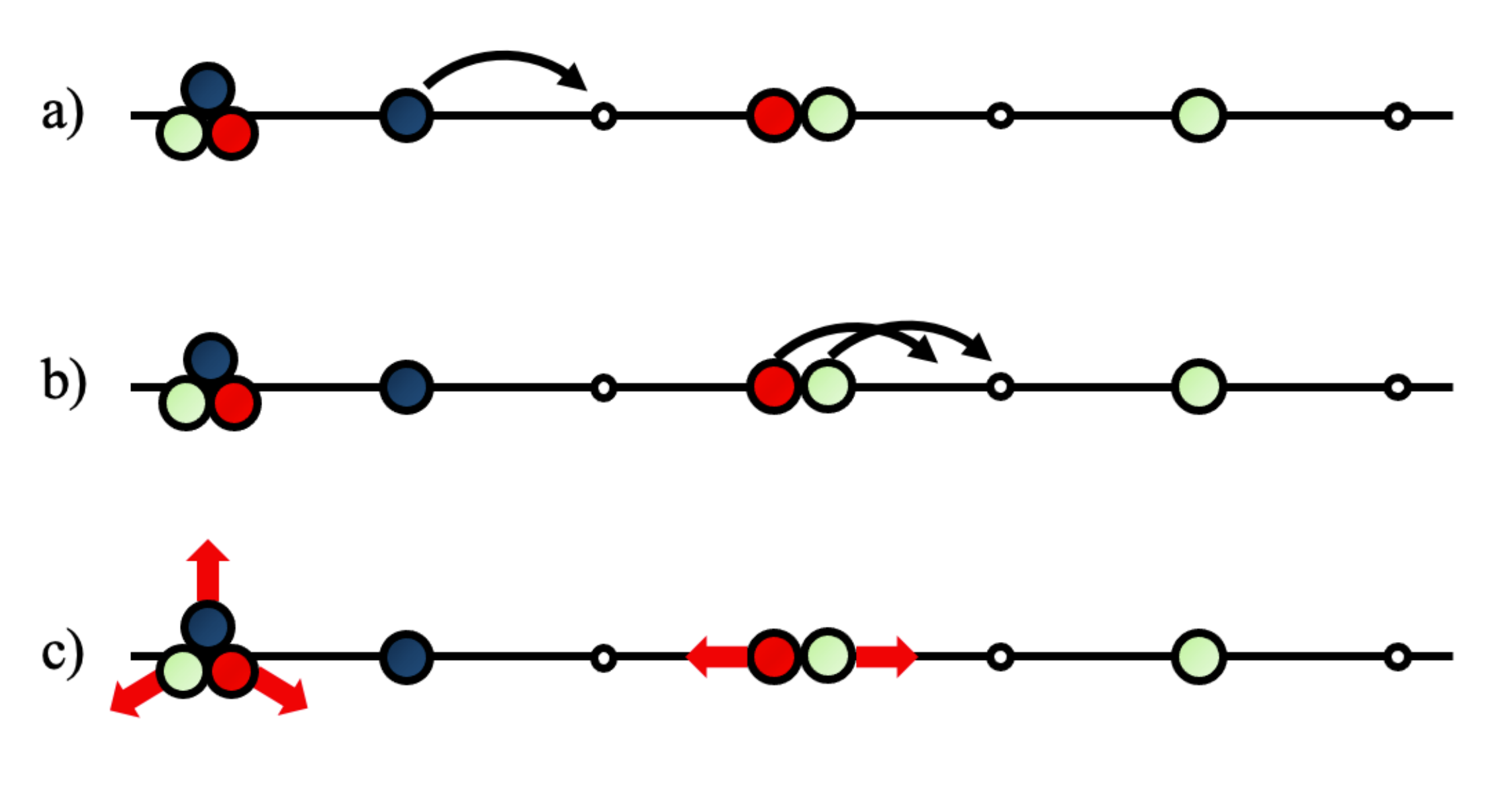}
  \caption{Schematics of individual terms in the Hamiltonian $H_{\rm OBC}$ (Eq. \eqref{eq:ham}). The case of $N=3$ is shown. (a)~the one-body hopping term $\hat{H}_1$, (b)~the ($N-1$)-body hopping term $\hat{H}_{N-1}$, and (c)~the on-site repulsive or attractive interaction term $\hat{H}_U$. Blue, red, green (black, gray, white) balls represent fermions with flavors $\sigma=1,2,3$.}
  \label{fig:schematic_Hamiltonian}
\end{figure}

A schematic of each term in the Hamiltonian is shown in Fig.~\ref{fig:schematic_Hamiltonian}. The first term $\hat{H}_1$ describes the one-body hopping term. The second term $\hat{H}_{N-1}$ represents the ($N-1$)-body-hopping term. The hopping amplitude $t \in \mathbb{R}$ is common to $\hat{H}_1$ and $\hat{H}_{N-1}$. The third term $\hat{H}_U$ represents on-site repulsive $(U>0)$ or attractive $(U<0)$ interactions. To simplify the following discussion, we include a uniform potential in the third term, which produces 
a constant shift 
in the energies of the eigenstates in each particle-number sector.

\medskip

\subsection{$\eta$-operator}
Next, we consider the $\eta$-operator. In the case of $N=2$, the $\eta$-operator is defined as $\hat{\tilde{\eta}}^\dagger=\sum_{x=1}^L e^{i\pi x}\hat{\eta}^{\dagger}_{x}$.
Naively, one might think that the same definition works for $N >2$. Indeed, when $N$ is even, we have a series of eigenstates created by applying $\hat{\tilde{\eta}}^\dagger$ to the vacuum state. However, this does not work when $N$ is odd, because $\hat{\tilde{\eta}}^\dagger$ squares to zero. To avoid this problem, we define the $\eta$-operator as
\begin{equation}
    \hat{\eta}^\dagger=\sum_{x=1}^L e^{i\pi x} \hat{U}_{1,\ldots,x-1} \hat{\eta}^{\dagger}_{x},
\end{equation}
where $\hat{U}_{1,\ldots,x-1}$ is a unitary operator defined as $\hat{U}_{1,\ldots,x-1}=e^{i \pi \sum_{j=1}^{x-1}
\hat{n}_j}$ for $x>1$ and $\hat{U}_{1,\ldots,x-1}=1$ for $x=1$.

\medskip

\subsection{Exact eigenstates of the Hamiltonian}

By repeatedly applying $\hat{\eta}^\dagger$ to the vacuum state, we have a series of states, namely, {\it $\eta$-clustering states}
\begin{equation}
\begin{split}
    \ket{\Phi^{L}_{M}}
    &:=\frac{1}{M!}(\hat{\eta}^\dagger)^M \ket{0} \\
  &=\sum_{1\leq x_1<\ldots <x_M \leq L} \, \left\{\prod_{j=1}^M e^{i\pi x_j} \hat{\eta}^{\dagger}_{x_j}\right\}\ket{0} \label{eq:Phi_{L,M}}.
\end{split}
\end{equation}

\begin{figure}
  \centering
  \includegraphics[width=1.0\linewidth]{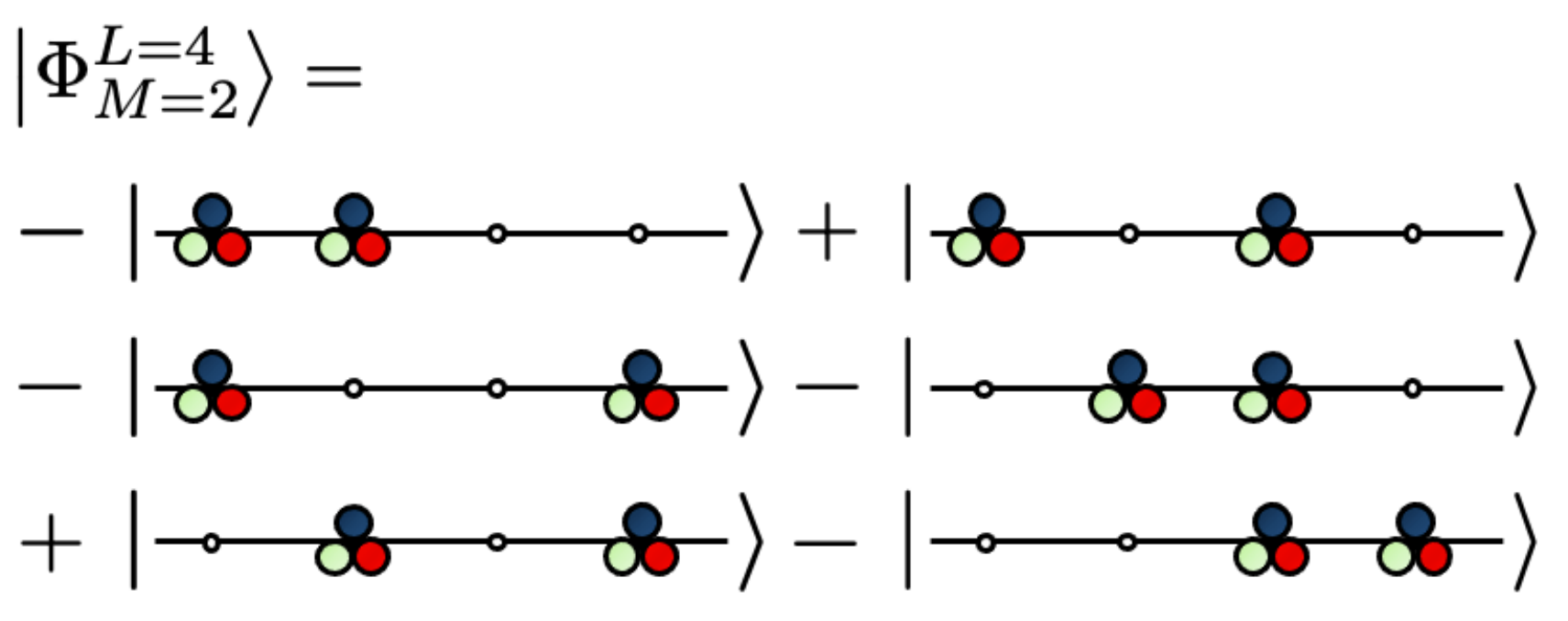}
  \caption{A Schematic of an $\eta$-clustering state $\ket{\Phi^{L}_{M}}$ for $N=3$, $L=4$, and $M=2$. The particle number of $\ket{\Phi^{L}_{M}}$ is $NM$.}
  \label{fig:philm}
\end{figure}
Figure~\ref{fig:philm} shows a schematic of $\eta$-clustering states. They do not vanish for $M=0,\ldots, L$, and 
boil down to $\eta$-pairing states $\frac{1}{M!}(\hat{\tilde{\eta}}^\dagger)^{M}\ket{0}$ when $N=2$. These state are not eigenstates of the original SU($N$) Hubbard model with the Hamiltonian $\hat{H}_1+\hat{H}_U$, but are exact eigenstates of $\hat{H}_\mathrm{OBC}=\hat{H}_1+\hat{H}_{N-1}+\hat{H}_U$.

{\it Theorem 1.}---For all $M=0,\ldots, L$, we have $\hat{H}_\mathrm{OBC}\ket{\Phi^{L}_{M}}=0$.

{\it Proof of Theorem 1.}---
First we define the following state:
\begin{equation}
  \ket{\Phi^{L}(\alpha)}=
  \left\{\prod_{x=1}^L (1+\alpha e^{i\pi x}\hat{\eta}^\dagger_x)\right\}\ket{0}=\sum_{M=0}^L \alpha^M \ket{\Phi^{L}_{M}},\label{eq:Phi}
\end{equation}
where $\alpha$ is a formal parameter. Since $\hat{H}_\mathrm{OBC}$ conserves the particle number, if $\hat{H}_\mathrm{OBC}\ket{\Phi^{L}(\alpha)}=0$, then $\hat{H}_\mathrm{OBC}\ket{\Phi^{L}_M}=0$ for all $M$. For notational simplicity, we write the hopping term between sites $x$ and $x+1$ as
\begin{align}
  \hat{T}_{x,{x+1}} &= \sum_{\sigma=1}^N \hat{T}^\sigma_{x,{x+1}}, \label{eq:htxy}\\
   \hat{T}^\sigma_{x,{x+1}} &=\left[(\hat{c}^\dagger_{x,\sigma} \hat{c}_{{x+1},\sigma}+\hat{\overline{c}}^\dagger_{x,\sigma} \hat{\overline{c}}_{{x+1},\sigma})+\text{h.c.}\right].
\end{align}
We also denote by $\mathcal{W}$ the subspace where the eigenvalue of $\hat{n}_x$ is $0$ or $N$ for all $x$. Then, one finds
\begin{equation}
  \begin{split}
& \left[ \hat{T}_{x,{x+1}}^{\sigma}, \hat{\eta}_{z}^\dagger \right] \mathcal{W} \\
&=\left(\delta_{x,z} +\delta_{{x+1},z}\right)  \left[\hat{c}^\dagger_{x,\sigma} \hat{\overline{c}}^\dagger_{{x+1},\sigma}+ \hat{c}^\dagger_{{x+1},\sigma} \hat{\overline{c}}^\dagger_{x,\sigma} \right] \mathcal{W},
\label{eq:commutation_relation_local}
\end{split}
\end{equation}
which yields
\begin{equation}
  \begin{split}
&\left[ \hat{T}_{x,{x+1}}^{\sigma}, (1+\alpha e^{i \pi x}\hat{\eta}_{x}^\dagger)(1- \alpha e^{i \pi x}\hat{\eta}_{x+1}^\dagger) \right]\mathcal{W} =0.
\label{eq:commutation_relation}
\end{split}
\end{equation}
Using this commutation relation and $\hat{T}_{x,{x+1}}^{\sigma}\ket{0}=0$, one can easily check that $(\hat{H}_1+\hat{H}_{N-1}) \ket{\Phi^{L}(\alpha)}=0$. From this and $\hat{H}_U \mathcal{W}=0$, we have $\hat{H}_\mathrm{OBC}\ket{\Phi^{L}(\alpha)}=0$, and therefore $\hat{H}_\mathrm{OBC}\ket{\Phi^{L}_M}=0$ for all $M$.
\hspace{\fill}$\blacksquare$

\medskip

{\it Remark 1.}---
In Theorem 1, we assumed open boundary conditions. Here, we consider the cases of (anti-)periodic boundary conditions. The Hamiltonian reads

\begin{align}
  \hat{H}_{\mathrm{(A)PBC}}=\hat{H}_\mathrm{OBC}-t\hat{T}^{\mathrm{(A)PBC}}_{L,1},  \label{eq:hambc}
\end{align}
where $\hat{T}^{\mathrm{(A)PBC}}_{L,1}:=\hat{T}_{L,L+1}$ and $\hat{c}_{L+1,\sigma}= \hat{c}_{1,\sigma}$ ($-\hat{c}_{1,\sigma}$) for periodic (anti-periodic) boundary conditions.

In these cases, $\eta$-clustering states are not necessarily eigenstates. The boundary conditions where $\eta$-clustering states are eigenstates are summarized in Table I (see Appendix~\ref{sec:pbc} for derivation). We also confirmed by exact diagonalization for small systems that the zero-energy eigenstates for each boundary condition are limited to them in the whole Fock space $\mathcal{V}$, except when the strength of the interaction is fine-tuned to the values
$U=0, \pm t$.

\begin{table}
  \label{tab:boundary_condition}
  \caption{The boundary conditions for which $\ket{\Phi^L_{M}}$ is an eigenstate with zero energy. We denote by O and (A)P the open boundary conditions and (anti-)periodic boundary conditions. The case of $M = 1,\ldots, L-1$ is shown. When $M=0,L$, $\ket{\Phi^L_{M}}$ is an eigenstate for any boundary conditions.}
  \centering
  \begin{tabular}{ccccc}
      \toprule
       & \multicolumn{2}{c}{ $\bm{N}$\textbf{: even} } & \multicolumn{2}{c}{ $\bm{N}$\textbf{: odd}}\\
       & $L$: even  & $L$: odd & $L$: even  & $L$: odd \\\cmidrule(lr){2-3}\cmidrule(lr){4-5}
      $M$: even & O, P, AP & O & O, AP & O, P \\[4pt]
      $M$: odd & O, P, AP & O & O, P & O, AP \\\bottomrule
  \end{tabular}
\end{table}

\medskip

{\it Remark 2.}---
In Theorem 1, for simplicity, we assumed that the Hamiltonian is SU($N$) symmetric and translationally invariant in the bulk. However, these two conditions are not necessary. To illustrate this, we consider the following Hamiltonian
\begin{equation}
\begin{split}
\hat{\tilde{H}}_\mathrm{OBC}&=-\sum_{x=1}^{L-1}\sum_{\sigma=1}^N t^{\sigma}_{x,x+1} \hat{T}^\sigma_{x,x+1} \\
&+\sum_{x=1}^{L}\sum_{\sigma,\tau=1}^N U^{\sigma,\tau}_x \left(\hat{n}_{x,\sigma}-\frac{1}{2}\right)\left(\hat{n}_{x,\tau}-\frac{1}{2}\right),
\label{eq:general_hamiltonian}
\end{split}
\end{equation}
where
$t_{x,x+1}^\sigma\in \mathbb{R}$ and $U^{\sigma,\tau}_x\in \mathbb{R}$ depend on $x,\sigma,\tau$.
Generally, this Hamiltonian is not SU($N$) symmetric nor translationally invariant \footnote{When $N=2$, the hopping term of \eqref{eq:general_hamiltonian} has the SU(2) symmetry even when $t_{x,x+1}^{\sigma=1}\neq t_{x,x+1}^{\sigma=2}$, because $\hat{T}^{\sigma=1}_{x,x+1}=\hat{T}^{\sigma=2}_{x,x+1}=\sum_{\sigma=1}^{2}(\hat{c}^\dagger_{x,\sigma} \hat{c}_{x+1,\sigma}+\text{h.c.})$ are SU($2$) symmetric by itself.
}. Then,
one can show that $\eta$-clustering states $\ket{\Phi^L_{M}}$ are eigenstates of $\hat{\tilde{H}}_\mathrm{OBC}$ for all $M=0,\ldots, L$. The proof goes along the same lines as the proof of Theorem 1.

\medskip

\section{Properties of $\eta$-clustering states}

\subsection{Entanglement entropy and the connection to quantum many-body scar states}
Here we compute the entanglement entropy of $\eta$-clustering states $\ket{\Phi^{L}_M}$ using the technique developed in Refs.~\cite{vafek_entanglement_2017,vedral_high-temperature_2004,fan_entanglement_2005}. We partition the $L$ sites $x=1,\ldots,L$ into a subsystem $A$ $(x=1,\ldots, L_A)$ and a subsystem $B$ $(x=L_A+1,\ldots, L)$ and compute the reduced density matrix by tracing out the degrees of freedom in
$B$. Then we compute the von Neumann entropy of the reduced density matrix, which we denote by $S_A$. This can be done for both odd and even $N$ in the same way as for the $\eta$-pairing states~\cite{vafek_entanglement_2017}. As a result, we obtain
\begin{equation}
    S_A= \frac{1}{2}(1+\ln [2\pi \nu (1-\nu)L_A]).
    \label{eq:entanglement_scar}
\end{equation}
Here, we take the thermodynamic limit $L,M\to \infty$ such that $\nu=M/L$ is kept constant. Equation~\eqref{eq:entanglement_scar} clearly shows that $S_A$
scales logarithmically with the system size, implying that $\eta$-clustering states $\ket{\Phi^{L}_M}$ have sub-volume-law entanglement, even though they are in the middle of the spectrum of $\hat{H}_\mathrm{OBC}$.

When $N=2$, $\eta$-pairing states are not examples of quantum many-body scars because the model has $\eta$-SU(2) symmetry and $\eta$-pairing states are the only eigenstates in their respective quantum number sectors~\cite{vafek_entanglement_2017, moudgalya_2020}. However, since the model does not have such symmetry for $N \ge 3$, $\eta$-clustering states may be regarded as quantum many-body scars.

\medskip

\subsection{Off-diagonal long-range order}
For a normalized state $\ket{\phi}$, we define the singlet correlation function by $\bra{\phi}\hat{\eta}^\dagger_x \hat{\eta}_y \ket{\phi}$. This is an extension of the pair correlation function. If this does not vanish when $|x-y|\to \infty$, we say that $\ket{\phi}$ exhibits $N$-particle ODLRO.
We write the singlet correlation function of $\eta$-clustering states as
\begin{equation}
      \langle \hat{\eta}^\dagger_x \hat{\eta}_y \rangle^L_M :=\frac{\bra{\Phi^L_{M}} \hat{\eta}^\dagger_x \hat{\eta}_y \ket{\Phi^{L}_{M}}}{\braket{\Phi^{L}_{M}}{\Phi^{L}_{M}}}.
\end{equation}
Then, $\langle \hat{\eta}^\dagger_x \hat{\eta}_y \rangle^L_M$ is calculated as follows.

{\it Theorem 2.}---
For $x\neq y$ and $0<M<L$, we have
\begin{equation}
    \langle \hat{\eta}^\dagger_x \hat{\eta}_y \rangle^L_M=(-1)^{r} \dfrac{M(L-M)}{L(L-1)}
    \label{eq:singlet_correlation_even}
\end{equation}
when $N$ is even and
\begin{equation}
    \langle \hat{\eta}^\dagger_x \hat{\eta}_y \rangle^L_M=\dfrac{\sum_{j=j_{\mathrm{min}}}^{j_{\mathrm{max}}} (-1)^{j}\binom{L-r-1}{j} \binom{r-1}{M-j-1}}{(-1)^{M+r-1} \binom{L}{M}}
    \label{eq:singlet_correlation_odd}
\end{equation}
when $N$ is odd, where $r=|x-y|$, $j_{\mathrm{min}}=\operatorname{max}\{0,M-r\}$, and ${j_{\mathrm{max}}}={\operatorname{min}\{L-r-1,M-1\}}$.

\medskip

\begin{figure}
  \centering
  \includegraphics[width=1.0\linewidth]{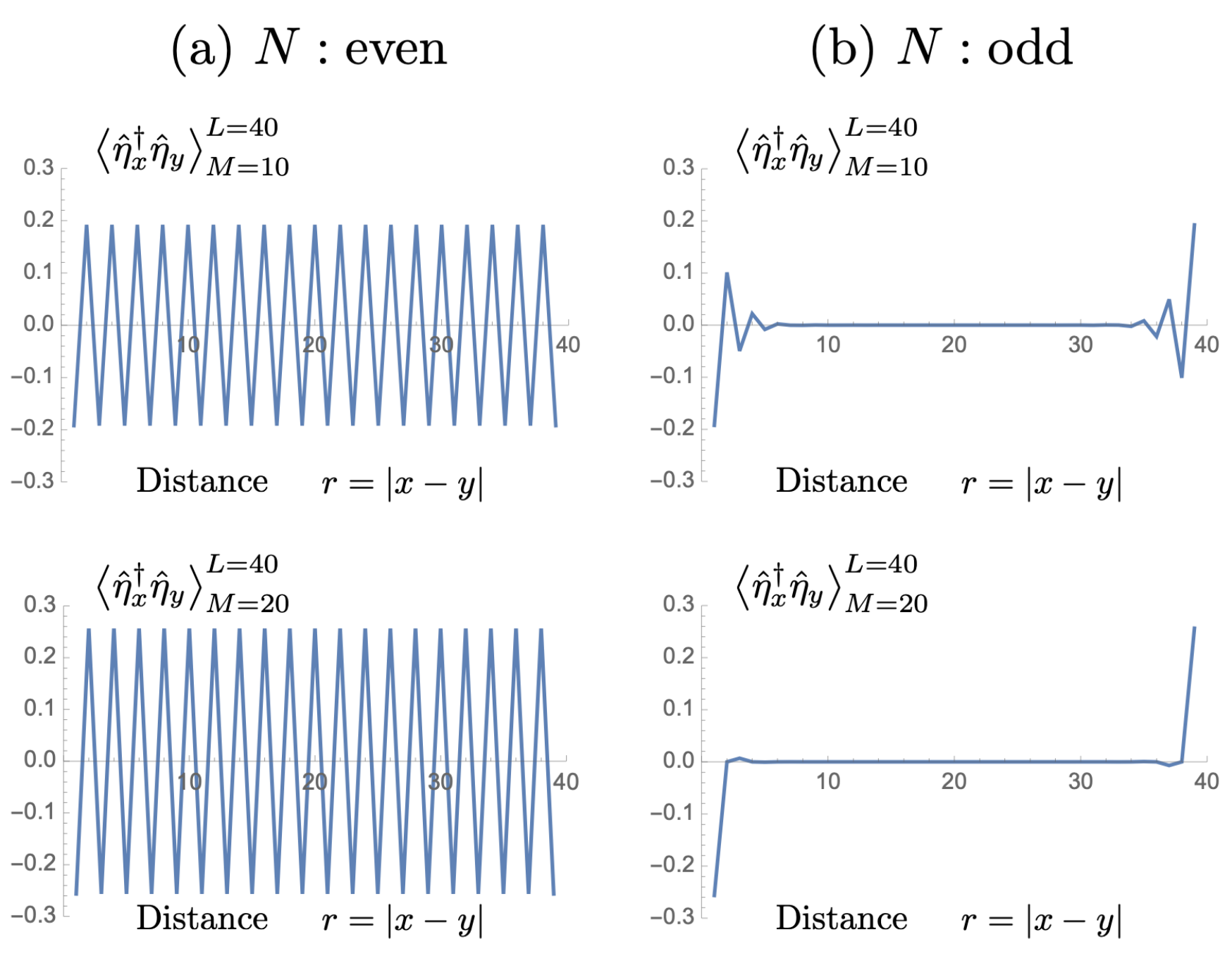}
  \caption{The singlet correlation function $\langle \hat{\eta}^\dagger_x \hat{\eta}_y \rangle^L_M$ as a function of $r=|x-y|$. We substituted $L=40$ and $M=10, 20$ for Eqs. \eqref{eq:singlet_correlation_even} and \eqref{eq:singlet_correlation_odd}. (a) When $N$ is even, the absolute value of the singlet correlation function does not depend on $r$. (b) When $N$ is odd, the singlet correlation function decays exponentially toward the other end of the chain, but shows a revival at $r \simeq L$.}
  \label{fig:odd_singlet_correlation_function}
\end{figure}

Figure~\ref{fig:odd_singlet_correlation_function} shows
$\langle \hat{\eta}^\dagger_x \hat{\eta}_y \rangle^L_M$ as a function of $r=|x-y|$ when $L=40$.
The result for even $N$ is consistent with the case of $N=2$~\cite{yang__1989}. In this case, from Theorem 2, 
$\eta$-clustering states have $N$-particle ODLRO in the limit where the filling $\nu=M/L$ is kept constant and $L\to \infty$. On the other hand, when $N$ is odd, the behavior of the singlet correlation function is very different. The singlet correlation function decays exponentially with distance
as $|\langle \hat{\eta}^\dagger_x \hat{\eta}_y \rangle^L_M| \sim e^{-r/\xi(L,M)}$ (see Appendix~\ref{sec:singlet_odd}), where $\xi(L,M)$ is the correlation length depending on $L$ and $M$.
However,
when $r \simeq L$, the singlet correlation function do not vanish.
This shows end-to-end long-range correlations. This behavior can be qualitatively understood by mapping our system to a spin chain (see Appendix~\ref{sec:singlet_spin}). In the limit $L\to \infty$, Eq.~\eqref{eq:singlet_correlation_odd} reads $\langle \hat{\eta}^\dagger_1 \hat{\eta}_L \rangle^L_M \to (-1)^{L+M}\nu(1-\nu)$. In the same limit, the correlation length is estimated as $\xi(L,M)\to -1/\log |(2\nu-1)|$. We note in passing that a similar behavior has been observed in a spinless fermion system~\cite{lang_topological_2015,iemini_majorana_2015}.

{\it Proof of Theorem 2.}---
We use $\ket{\Phi^{L}(\alpha)}$ defined by \eqref{eq:Phi} for the calculation. By direct calculation (see Appendix~\ref{sec:derivation_19}), we have
\begin{align}
  \braket{\Phi^{L}(\alpha)}&=(1+\alpha^2)^L
  \label{eq:norm_Phi_alpha}
\end{align}
and
\begin{equation}
  \begin{split}
&(-1)^{x+y}\left\langle\Phi^{L}(\alpha)\left|\hat{\eta}_{x}^{\dagger} \hat{\eta}_y\right| \Phi^{L}(\alpha)\right\rangle \\
&=
\begin{cases}
   \alpha^{2}\left(1+\alpha^{2}\right)^{L-2} &  N :\text {even},\\
  \alpha^{2}\left(1+\alpha^{2}\right)^{L-r-1}\left(1-\alpha^{2}\right)^{r-1} & N :\text {odd}.
\end{cases}
\end{split}
\label{eq:singlet_correlation_Phi_alpha}
\end{equation}
Since $\hat{\eta}_{x}^{\dagger} \hat{\eta}_y$ conserves the particle number, $\braket{\Phi^{L}_{M}}$ $\left(\bra{\Phi^L_{M}} \hat{\eta}^\dagger_x \hat{\eta}_y \ket{\Phi^{L}_{M}}\right)$ is the coefficient of $\alpha^{2M}$ in $\braket{\Phi^{L}(\alpha)}$ $\left(\bra{\Phi^{L}(\alpha)} \hat{\eta}^\dagger_x \hat{\eta}_y \ket{\Phi^{L}(\alpha)}\right)$. 
This yields the desired Eqs. \eqref{eq:singlet_correlation_even} and \eqref{eq:singlet_correlation_odd}.
\hspace{\fill}$\blacksquare$

Finally, we see the connection between the singlet correlation function and the $n$-particle reduced density matrix~\cite{yang_concept_1962}. We consider a state described by a density matrix $\hat{\rho}$. The $n$-particle reduced density matrix $\hat{\rho}_n$ is a matrix of size $(L\times N)^n$ with the matrix elements
\begin{equation}
\begin{split}
    &(\hat{\rho}_n)_{[(x_1,\sigma_1),\ldots, (x_n,\sigma_n)],[(y_1,\tau_1),\ldots, (y_n,\tau_n)]}\\
    &:= \Tr [\hat{c}^\dagger_{x_n,\sigma_n} \ldots \hat{c}^\dagger_{x_1,\sigma_1} \hat{c}_{y_1,\tau_1}\ldots \hat{c}_{y_n,\tau_n} \hat{\rho}]
\end{split}
\end{equation}
where $[(x_1,\sigma_1),\ldots, (x_n,\sigma_n)]$ and $[(y_1,\tau_1),\ldots, (y_n,\tau_n)]$ $(x_i,y_i=1,\ldots, L,\ \sigma_i,\tau_i=1,\ldots, N)$ denote sequences of sites and flavors. We write the maximum eigenvalue of $\hat{\rho}_n$ as $\lambda_n$. Then, another characterization of $n$-particle ODLRO is that $\lambda_n=O(L)$. The singlet correlation functions are off-diagonal elements of $\hat{\rho}_n$ where $x_1=\ldots =x_N$ and $y_1=\ldots =y_N$. If we consider $\ket{\Phi^{L}_{M}}$, the remaining off-diagonal elements vanish. Thus we have
\begin{numcases}{\lambda_N=}
O(L) &  $N$: \text {even},\label{eq:max_eigenvalue_even}\\
O(1) & $N$: \text {odd}\label{eq:max_eigenvalue_odd},
\end{numcases}
and
\begin{equation}
    \lambda_n =0\quad \text{when}\quad n<N.
\end{equation}

In Ref.~\cite{yang_concept_1962}, Yang conjectured that there exists a numerical constant $\beta_n$ such that
\begin{numcases}{\lambda_n\leq}
\beta_n L^{n/2} &  $N$: \text {even},\label{eq:max_eigenvalue_bound_even}\\
\beta_n L^{(n-1)/2} & $N$: \text {odd}\label{eq:max_eigenvalue_bound_odd}.
\end{numcases}
If we accept the conjecture, $\eta$-clustering states do not saturate the upper bound when $N\geq 3$, while in the case of $N=2$, it is shown that $\eta$-pairing states maximize $\lambda_2$~\cite{nakagawa_eta_2021}.

\medskip

\section{Parent Hamiltonian for $\eta$-clustering states}
\label{sec:parent}
In the SU(2) Hubbard model, $\eta$-pairing states are not ground states. On the other hand, adding terms to the Hamiltonian allows us to create models
whose ground states are $\eta$-pairing
states~\cite{essler_new_1992, essler_electronic_1993, arrachea_exact_1994, de_boer_1995, schadschneider_superconductivity_1995, de_boer_exact_1995}. In light of these contexts, we seek a parent Hamiltonian for which $\eta$-clustering states $\ket{\Phi^{L}_{M}}\ (M=0,\ldots,L)$ are the {\it unique} ground states. For this purpose, consider the following 
Hamiltonian consisting of the $N$-body hopping, the nearest-neighbor interaction, and the on-site potential terms \footnote{By considering this term, we can also construct an SU(3) symmetric model where both the $\eta$-pairing and the three-body $\eta$-clustering states are exact energy-eigenstates (see Appendix~\ref{sec:coexist})}
:
\begin{equation}
  \hat{H}_{V} = V\sum_{x=1}^{L-1} \hat{V}_{x,x+1},
  \label{eq:hv}
\end{equation}
where
\begin{equation}
  \begin{split}
    \hat{V}_{x,x+1} =& \frac{1}{2}(\hat{\eta}^{\dagger}_{x}\hat{\eta}_{x+1}+\text{h.c.)}
    \\ &-\frac{1}{N^2} \hat{n}_x \hat{n}_{x+1}+\frac{1}{2N} (\hat{n}_x+\hat{n}_{x+1}).
  \end{split}
\end{equation}

Then we prove the following theorem.

{\it Theorem 3.}---
Consider the Hamiltonian $\hat{H}_\mathrm{OBC}^{\prime}=\hat{H}_\mathrm{OBC}+\hat{H}_{V}$. If $U\leq 0$ and $V> 8N^2|t|$, then the zero-energy ground states of $\hat{H}_\mathrm{OBC}^{\prime}$ in the whole Fock space $\mathcal{V}$ are exactly $(L+1)$-fold degenerate and written as $\ket{\Phi^{L}_{M}} (M=0,\ldots, L)$.

\medskip

To prove Theorem 3, we consider the local Hamiltonian $\hat{H}_{x,x+1}= t \hat{T}_{x,x+1}+V \hat{V}_{x,x+1}$, where $\hat{T}_{x,x+1}$ is defined as Eq.~\eqref{eq:htxy}. We denote by $\mathcal{V}_{x,x+1}$ the $2^{2N}$-dimensional subspace of $\mathcal{V}$ spanned by states of the form $\{ \prod_{\sigma=1}^N (\hat{c}^\dagger_{x,\sigma})^{n_{x,\sigma}} (\hat{c}^\dagger_{x+1,\sigma})^{n_{x+1,\sigma}} \} \ket{0}$ with $n_{x,\sigma}, n_{x+1,\sigma}=0,1$.

{\it Lemma 4.}---
If $V> 8N^2|t|$, the zero-energy ground states of $\hat{H}_{x,x+1}$ in the subspace $\mathcal{V}_{x,x+1}$ are threefold degenerate and written as $\ket{0}$, $\hat{\eta}^\dagger_x\hat{\eta}^\dagger_{x+1}\ket{0}$, and $(\hat{\eta}^\dagger_x-\hat{\eta}^\dagger_{x+1})\ket{0}$.

{\it Proof of Lemma 4}---
To obtain the ground state of $\hat{V}_{x,x+1}$, we map the $\eta$-operators to spin-$1/2$ operators $\{\hat{S}^+_x, \hat{S}^-_x, \hat{S}_x^z\}$ defined on the one-dimensional lattice with $L$ sites. Let $\hat{P}$ be a projector to the subspace $\mathcal{W}$. Then we define
\begin{equation}
  \begin{split}
     \hat{S}_x^+ &:=e^{i\pi x} \hat{P} \hat{U}_{1,\ldots,x-1} \hat{\eta}_{x}^{\dagger} \hat{P}, \\
     \hat{S}_x^- &:=e^{i\pi x} \hat{P} \hat{U}_{1,\ldots,x-1} \hat{\eta}_x \hat{P}, \\
     \hat{S}_x^z &:=\hat{P} \left(\hat{\eta}_{x}^{\dagger} \hat{\eta}_x-\frac{1}{2}\right) \hat{P}.
     \label{eq:spin_map}
  \end{split}
\end{equation}
The operators $\{\hat{S}^+_x, \hat{S}^-_x, \hat{S}_x^z\}$ satisfy the usual commutation relations $\left[\hat{S}^+_j,\hat{S}^-_k\right]=2\delta_{j,k} \hat{S}^z_j$ and $\left[\hat{S}^z_j,\hat{S}^\pm_k\right]=\delta_{j,k} \hat{S}^\pm_j$.

Let us express $\hat{V}_{x,x+1}$ with the spin operators. We write $\hat{Q}=1-\hat{P}$. Since $[\hat{V}_{x,x+1},\hat{P}]=0$, it is expressed as
\begin{equation}
  \begin{split}
      \hat{V}_{x,x+1}
      &= \hat{P}\hat{V}_{x,x+1}\hat{P}+\hat{Q}\hat{V}_{x,x+1}\hat{Q} \\
      &= \hat{P} \left(-\hat{\bm{S}}_x\cdot \hat{\bm{S}}_{x+1}+\frac{1}{4}\right) \hat{P} \\
      &+\hat{Q}\left[\frac{1}{4}- \left(\frac{\hat{n}_x}{N}-\frac{1}{2}\right)\left(\frac{\hat{n}_{x+1}}{N}-\frac{1}{2}\right)\right]\hat{Q}.
  \end{split}
\end{equation}
Let us consider the subspace $\mathcal{V}_{x,x+1}$. One finds that the first term acts nontrivially on $\mathcal{V}_{x,x+1}\cap \mathcal{W}$ and the second term acts nontrivially on $\mathcal{V}_{x,x+1}\cap \mathcal{W}^\bot$, where $\mathcal{W}^{\bot}$ is the orthogonal complement of $\mathcal{W}$. In the subspace $\mathcal{V}_{x,x+1}\cap \mathcal{W}$, the energy of the ground states is $0$ (the spin triplet), while that of the excited state is $1$ (the spin singlet). The corresponding ground states can be written as $\ket{0}$, $\hat{\eta}^\dagger_x\hat{\eta}^\dagger_{x+1}\ket{0}$, and $(\hat{\eta}^\dagger_x-\hat{\eta}^\dagger_{x+1})\ket{0}$. On the other hand, in the subspace $\mathcal{V}_{x,x+1}\cap W^\bot$, the energy of the ground state is $1/2N$. Therefore, the ground states of $\hat{V}_{x,x+1}$ in $\mathcal{V}_{x,x+1}$ are $\ket{0}$, $\hat{\eta}^\dagger_x\hat{\eta}^\dagger_{x+1}\ket{0}$, and $(\hat{\eta}^\dagger_x-\hat{\eta}^\dagger_{x+1})\ket{0}$ and their energy is 0. The energy of the first excited state is $1/2N$.

In order to investigate the conditions for the ground states of $\hat{V}_{x,x+1}$ to be the unique ground states of $\hat{H}_{x,x+1}$, we use Weyl's theorem \cite{franklin_matrix_2012}. Writing the $k$-th eigenvalue of a matrix $A$ from the lowest by $\mu_k (A)$ and the operator norm of $A$ by $\|A\|$, we have 
\begin{equation}
  \mu_{k}\left(V \hat{V}_{x,x+1}\right)-\left\|t \hat{T}_{x,{x+1}}\right\| \leq \mu_{k}\left(\hat{H}_{x,x+1}\right).
  \label{eq:Weil_two}
\end{equation}
Since the energy of the first excited state of $\hat{V}_{x,x+1}$ is $1/2N$, we have $\mu_{4} (\hat{V}_{x,x+1})=1/2N$. One can also evaluate $\|\hat{T}_{x,{x+1}}\|$ as
\begin{equation}
\begin{split}
        \|\hat{T}_{x,{x+1}}\|
        &= \left\| \sum_{\sigma=1}^N \left[(\hat{c}^\dagger_{x,\sigma} \hat{c}_{{x+1},\sigma}+\hat{\overline{c}}^\dagger_{x,\sigma} \hat{\overline{c}}_{{x+1},\sigma})+\text{h.c.}\right] \right\| \\
        &\leq2N \left(\left\| \hat{c}^\dagger_{x,\sigma}\right\|\left\| \hat{c}_{{x+1},\sigma}\right\|
        +\left\|\hat{\overline{c}}^\dagger_{x,\sigma} \right\|\left\|\hat{\overline{c}}_{{x+1},\sigma}\right\|\right)\leq 4N.
\end{split}
\end{equation}
Substituting them to Eq.~\eqref{eq:Weil_two}, we have
\begin{equation}
  \frac{V}{2N}-4N|t| \leq \mu_{4}\left(\hat{H}_{x,x+1}\right).
  \label{eq:fourth_eigenvalue}
\end{equation}
From \eqref{eq:fourth_eigenvalue}, one finds $\mu_{4}\left(\hat{H}_{x,x+1}\right)>0$ when $V> 8N^2|t|$. Since the threefold ground states of $\hat{V}_{x,x+1}$ are zero-energy eigenstates of $\hat{H}_{x,x+1}$, they are the unique ground states of $\hat{H}_{x,x+1}$.
\hspace{\fill}$\blacksquare$

Theorem 3 follows from Lemma 4.

{\it Proof of Theorem 3}---
First, $\hat{H}_\mathrm{OBC}^{\prime}$ can be written as
\begin{equation}
  \begin{split}
  \hat{H}_\mathrm{OBC}^{\prime}&= \sum_{x=1}^{L-1} \hat{H}_{x,x+1} +\hat{H}_U.
  \label{eq:Hamiltonian_decomposition}
\end{split}
\end{equation}
It follows from Lemma 4 that, if $V> 8N^2|t|$, $\hat{H}_{x,x+1}$ is positive semidefinite in the subspace $\mathcal{V}_{x,x+1}$. Since $\hat{H}_{x,x+1}$ acts nontrivially only on the sites $x$ and $x+1$, $\hat{H}_{x,x+1}$ is still positive semidefinite in the whole Fock space $\mathcal{V}$. Since $U\sum_{x=1}^{L} \hat{n}_x\left(\hat{n}_x-N\right)$ is also positive semidefinite when $U\leq 0$, $\hat{H}_\mathrm{OBC}^{\prime}$ is positive semidefinite. Thus, any eigenstate of $\hat{H}_\mathrm{OBC}^{\prime}$ with zero eigenvalue is a ground state.
From Lemma 4, we see that $\hat{H}_{x,x+1}\ket{\Phi^{L}_{M}}=0$ for all $x$. This, together with $\hat{H}_{U}\ket{\Phi^{L}_{M}}=0$, implies that $\hat{H}_\mathrm{OBC}^{\prime}\ket{\Phi^{L}_{M}}=0$. Therefore, $\ket{\Phi^{L}_{M}}$ are ground states of $\hat{H}_\mathrm{OBC}^{\prime}$.

Next, we show that $\ket{\Phi^{L}_{M}}$ are the only ground states. We denote by $\hat{P}_{x,{x+1}}^{\mathrm{GS}}$ a projector to the (highly degenerate) ground states of $\hat{H}_{x,x+1}$ in the whole Fock space $\mathcal{V}$. Then, $[\hat{P},\hat{P}_{x,{x+1}}^{\mathrm{GS}}]=0$. From Lemma 4, we see that $\hat{H}_{x,x+1}$ can be expressed as
\begin{equation}
\begin{split}
     \hat{H}_{x,x+1}
     &=\hat{P} \Delta E\left(1-\hat{P}_{x,{x+1}}^{\mathrm{GS}}\right)\hat{P} \\
     &+\hat{Q}\Delta E \left(1-\hat{P}_{x,{x+1}}^{\mathrm{GS}}\right)\hat{Q}+\hat{D}_{x,x+1}  \\
     &= \hat{P}  \left[\Delta E\left(-\hat{\bm{S}}_x\cdot \hat{\bm{S}}_{x+1}+\frac{1}{4}\right)\right] \hat{P} \\
     &+ \hat{Q} \left[\Delta E  \left(1-\hat{P}_{x,{x+1}}^{\mathrm{GS}}\right)\right] \hat{Q}+\hat{D}_{x,x+1},
\end{split}
\end{equation}
  where we denote by $\Delta E(>0)$ the energy gap between the ground state and the first excited state of $\hat{H}_{x,x+1}$, and by $\hat{D}_{x,x+1}$ a positive semidefinite operator. Substituting this into Eq. \eqref{eq:Hamiltonian_decomposition} and setting $\hat{D}=\sum_{x=1}^{L-1} \hat{D}_{x,x+1}+\hat{H}_U$, we have
\begin{equation}
  \begin{split}
  \hat{H}_\mathrm{OBC}^{\prime}
  &=  \hat{P}\left[\Delta E \sum_{x=1}^{L-1} \left(-\hat{\bm{S}}_x\cdot \hat{\bm{S}}_{x+1}+\frac{1}{4}\right)\right]\hat{P} \\
  &+ Q\left[\Delta E \sum_{x=1}^{L-1}\left(1-\hat{P}_{x,x+1}^{\mathrm{GS}}\right)\right]Q +\hat{D} \\
  &= \hat{H}_\mathrm{OBC}^{\prime\prime}+\hat{D},
\end{split}
\end{equation}
where we introduced the notation $\hat{H}_\mathrm{OBC}^{\prime\prime}:=\hat{H}_\mathrm{OBC}^{\prime}-\hat{D}$. First, we consider the Hamiltonian $\hat{H}_\mathrm{OBC}^{\prime\prime}$. We see that the first (second) term acts nontrivially on $\mathcal{W}$ $(\mathcal{W}^\bot)$. The first term is the ferromagnetic Heisenberg model, so the ground states are $(L+1)$-fold degenerate in the subspace $\mathcal{W}$ and the energy of the ground states is $0$~\cite{tasaki2020}. For the second term, $\hat{P}_{x,x+1}^{\mathrm{GS}}\ket{\phi}=0$ for some $x$ if $\ket{\phi}\in\mathcal{W}^\bot$, so the eigenvalues of $\hat{H}_\mathrm{OBC}^{\prime\prime}$ in the subspace $\mathcal{W}^\bot$ are greater than or equal to $\Delta E$. Therefore, in the whole space $\mathcal{V}$, the ground states of $\hat{H}_\mathrm{OBC}^{\prime\prime}$ are $(L+1)$-fold degenerate and the energy is $0$. Next, we consider the effect of $\hat{D}$. Since $\hat{D}$ is positive semidefinite, the degeneracy of zero-energy eigenstates of $\hat{H}_\mathrm{OBC}^{\prime}=\hat{H}_\mathrm{OBC}^{\prime\prime}+\hat{D}$ is equal to or smaller than $L+1$. This, together with the fact that $\ket{\Phi^{L}_{M}}$ are zero-energy ground states of $\hat{H}_\mathrm{OBC}^{\prime}$, implies that they are the unique ground states of $\hat{H}_\mathrm{OBC}^{\prime}$ in the whole Fock space $\mathcal{V}$.
\hspace{\fill}$\blacksquare$

\section{Conclusions and outlook}
We have presented $N$-particle generalizations of $\eta$-pairing states in a chain of $N$-component fermions and constructed a model in which these states are exact eigenstates of the Hamiltonian for arbitrary $N$. When $N$ is even, these states exhibit $N$-particle ODLRO. Thus they serve as examples of multi-particle clustering of lattice fermions~\cite{schlottmann_1994,wu_hidden_2006}, which is relevant to charge 4e superconductors~\cite{kivelson_doped_1990,wu_hidden_2006, berg_charge-4e_2009}. When $N$ is odd, long-range correlation is absent inside the bulk, but there exists an $N$-particle long-range edge correlation. We have also constructed a model, 
whose unique ground states are the generalizations of $\eta$-pairing states.
For $N$ even,
the results can be generalized to any bipartite lattice in any dimension.
In the future, it would be interesting to consider
possible realizations of these states
with ultracold atoms in optical lattices.

\begin{acknowledgments}
We thank Masaya Nakagawa for valuable discussions.
H.K. was supported by JSPS Grant- in-Aid for Scientific Research on Innovative Areas: No. JP20H04630, JSPS KAKENHI Grant No. JP18K03445, No. JP21H05191, and the Inamori Foundation. H.Y. acknowledges the support of the Forefront Physics and Mathematics Program to Drive Transformation.
\end{acknowledgments}

\medskip

\appendix

\section{ZERO-ENERGY EIGENSTATES WITH (ANTI-)PERIODIC BOUNDARY CONDITIONS}
\label{sec:pbc}
Since $\ket{\Phi^L_M}$ is a zero-energy eigenstate of $\hat{H}_\mathrm{OBC}$, it is a zero-energy eigenstate of $\hat{H}_{\mathrm{(A)PBC}}=\hat{H}_\mathrm{OBC}- t\hat{T}^{\mathrm{(A)PBC}}_{L,1}$ if and only if it is a zero-energy eigenstate of $\hat{T}^{\mathrm{(A)PBC}}_{L,1}$. When $M=0$ or $L$, it is obvious that $\ket{\Phi^L_M}$ is a zero-energy eigenstate of $\hat{T}^{\mathrm{(A)PBC}}_{L,1}$. In the following, we consider the case of $M=1,\ldots, L-1$.

For notational simplicity, we write $\ket{\Phi^L(\alpha)}$ as
\begin{equation}
    \ket{\Phi^L(\alpha)}=\left\{
    \prod_{x=1}^L {\hat A}^\dagger_x (\alpha)
    \right\} \ket{0},
\end{equation}
where
\begin{equation}
    {\hat A}^\dagger_x (\alpha):= 1+ \alpha e^{i\pi x}\hat{\eta}^\dagger_x.
\end{equation}
When $N$ is even, each ${\hat A}^\dagger_x (\alpha)$ commutes with each other. Thus we have
\begin{equation}
    \begin{split}
    \hat{T}^{\mathrm{PBC}}_{L,1} \ket{\Phi^L(\alpha)}
    &= \hat{T}^{\mathrm{PBC}}_{L,1} \left\{
    \prod_{x=1}^L {\hat A}^\dagger_x (\alpha)
    \right\} \ket{0}
     \\
    &= \hat{T}^{\mathrm{PBC}}_{L,1}
    {\hat A}^\dagger_L (\alpha)
    {\hat A}^\dagger_1 (\alpha)
    \left\{
    \prod_{x=2}^{L-1} {\hat A}^\dagger_x (\alpha)
    \right\} \ket{0}
    \\
    &= \left[\hat{T}^{\mathrm{PBC}}_{L,1}, {\hat A}^\dagger_L (\alpha)
    {\hat A}^\dagger_1 (\alpha)
    \right]
    \left\{
    \prod_{x=2}^{L-1} {\hat A}^\dagger_x (\alpha)
    \right\} \ket{0}.
    \end{split}
\end{equation}
Note that $\hat{T}^{\mathrm{PBC}}_{L,1}=\hat{T}_{L,1}$. Using the commutation relation \eqref{eq:commutation_relation_local},
\begin{equation}
\begin{split}
&\left[\hat{T}^{\mathrm{PBC}}_{L,1}, {\hat A}^\dagger_L (\alpha)
    {\hat A}^\dagger_1 (\alpha)\right]\mathcal{W} \\
    &=\begin{cases}
   0 &  L :\text {even },\\
   -2 \alpha \sum_{\sigma=1}^N \left[\hat{c}^\dagger_{L,\sigma} \hat{\overline{c}}^\dagger_{1,\sigma}+ \hat{c}^\dagger_{1,\sigma} \hat{\overline{c}}^\dagger_{L,\sigma} \right]\mathcal{W} & L :\text {odd }.
\end{cases}
\end{split}
\end{equation}
Finally, when $N$ is even, $\hat{T}^{\mathrm{APBC}}_{L,1}=-\hat{T}^{\mathrm{PBC}}_{L,1}$. Therefore, if $L$ is even, $\ket{\Phi^L_M}$ is a zero-energy eigenstate of $\hat{T}^{\mathrm{(A)PBC}}_{L,1}$ for all $M=1,\ldots, L-1$. On the other hand, if $L$ is odd, $\ket{\Phi^L_M}$ is not an eigenstate of $\hat{T}^{\mathrm{(A)PBC}}_{L,1}$ for any $M=1,\ldots, L-1$. Here we used that $\hat{T}^{\mathrm{(A)PBC}}_{L,1}$ conserves the particle number.
When $N$ is odd, we consider the following states (see Ref.~\cite{kawabata_exact_2017} for a similar argument).

\begin{align}
  \ket{\Phi^L_+ (\alpha)}&:=\ket{\Phi^L(\alpha)}+\ket{\Phi^L(-\alpha)}\nonumber \\
  &=2\sum_{M:\text{ even}}\alpha^M \ket{\Phi^{L}_{M}}, \\
  \ket{\Phi^L_-(\alpha)}&:=\ket{\Phi^L(\alpha)}-\ket{\Phi^L(-\alpha)}\nonumber \\
  &=2\sum_{M:\text{ odd}}\alpha^M \ket{\Phi^{L}_{M}}.
\end{align}
Since
\begin{equation}
    \begin{split}
        &\ket{\Phi^L(\pm \alpha)} \\
        &=\left\{\prod_{x=1}^{L} {\hat A}^\dagger_x (\pm\alpha)\right\}\ket{0}
        =\left\{\prod_{x=1}^{L-1} {\hat A}^\dagger_x (\pm\alpha)\right\}{\hat A}^\dagger_L (\pm\alpha)\ket{0}\\
        &=\left\{\prod_{x=1}^{L-1} {\hat A}^\dagger_x (\pm\alpha)\right\}\ket{0}\pm \alpha e^{i \pi
        L}\hat{\eta}^{\dagger}_L\left\{\prod_{x=1}^{L-1} {\hat A}^\dagger_x (\mp\alpha)\right\}\ket{0},
    \end{split}
\end{equation}
one finds
\begin{align}
      &\ket{\Phi^L_\pm (\alpha)} = \ket{\Phi^L(\alpha)}\pm\ket{\Phi^L(-\alpha)}\nonumber\\
      &=  {\hat A}^\dagger_L(\mp\alpha)\prod_{x=1}^{L-1} {\hat A}^\dagger_x (\alpha)\ket{0}\pm{\hat A}^\dagger_L(\pm\alpha)\prod_{x=1}^{L-1} {\hat A}^\dagger_x (-\alpha)\ket{0}\nonumber \\
      &= {\hat A}^\dagger_L(\mp\alpha){\hat A}^\dagger_1(\alpha)\prod_{x=2}^{L-1} {\hat A}^\dagger_x (\alpha)\ket{0}\nonumber \\
      &\pm{\hat A}^\dagger_L(\pm\alpha){\hat A}^\dagger_1(-\alpha)\prod_{x=2}^{L-1} {\hat A}^\dagger_x (-\alpha)\ket{0}.
\end{align}
Since $\hat{T}^{\mathrm{(A)PBC}}_{L,1}$ act nontrivially only on the sites $x=L$ and $1$, they commute with $\prod_{x=2}^{L-1}  {\hat A}^\dagger_x (\pm\alpha)$. Finally, using the commutation relation \eqref{eq:commutation_relation_local}, we have
\begin{align}
    [\hat{T}^{\mathrm{PBC}}_{L,1},{\hat A}^\dagger_L( \pm e^{i \pi L} \alpha){\hat A}^\dagger_1( \pm \alpha)]\mathcal{W}&=0, \\
    [\hat{T}^{\mathrm{APBC}}_{L,1},{\hat A}^\dagger_L( \pm e^{i \pi L} \alpha){\hat A}^\dagger_1( \mp \alpha)]\mathcal{W}
    &=0.
\end{align}
Thus $\ket{\Phi^L_{+(-)} (\alpha)}$ is a zero-energy eigenstate of $\hat{T}^{\mathrm{PBC}}_{L,1}$ when $L$ is odd (even) and of $\hat{T}^{\mathrm{APBC}}_{L,1}$ when $L$ is even (odd). Otherwise it is not an eigenstate of $\hat{T}^{\mathrm{(A)PBC}}_{L,1}$ because
\begin{align}
    &[\hat{T}^{\mathrm{PBC}}_{L,1},{\hat A}^\dagger_L( \pm e^{i \pi L} \alpha){\hat A}^\dagger_1( \mp \alpha)]\mathcal{W} \nonumber\\
    &=\pm2 \alpha \sum_{\sigma=1}^N  \left[\hat{c}^\dagger_{L,\sigma} \hat{\overline{c}}^\dagger_{1,\sigma}+ \hat{c}^\dagger_{1,\sigma} \hat{\overline{c}}^\dagger_{L,\sigma} \right]\mathcal{W}, \\
    &[\hat{T}^{\mathrm{APBC}}_{L,1},{\hat A}^\dagger_L( \pm e^{i \pi L} \alpha){\hat A}^\dagger_1( \pm \alpha)]\mathcal{W} \nonumber\\
    &=\pm 2 \alpha \sum_{\sigma=1}^N \left[\hat{c}^\dagger_{L,\sigma} \hat{\overline{c}}^\dagger_{1,\sigma}- \hat{c}^\dagger_{1,\sigma} \hat{\overline{c}}^\dagger_{L,\sigma} \right]\mathcal{W}.
\end{align}
Therefore, we have Table I.

\medskip

\medskip

\begin{widetext}
\section{EVALUATION OF THE SINGLET CORRELATION FUNCTION AT EDGES WHEN $N$ IS ODD}
\label{sec:singlet_odd}
We evaluate the singlet correlation \eqref{eq:singlet_correlation_odd} when $r$ is small. When $r\leq M,L-M$, one finds $j_{\mathrm{min}}=\operatorname{max}\{0,M-r\}=M-r$ and ${j_{\mathrm{max}}}={\operatorname{min}\{L-r-1,M-1\}}=M-1$. Thus we have

\begin{equation}
\begin{split}
    |\langle \hat{\eta}^\dagger_x \hat{\eta}_y \rangle^L_M|\times \binom{L}{M}
    &= \left|\sum_{j=M-r}^{M-1} \binom{L-r-1}{j} (-1)^j \binom{r-1}{M-j-1}\right| \\
    &= \left|\sum_{j=0}^{r-1} \binom{L-r-1}{M-r+j} (-1)^j \binom{r-1}{j}\right|. \\
    \label{eq:singlet_approx}
\end{split}
\end{equation}
Now we consider a sequence of functions that satisfies the following recurrence relation.
\begin{equation}
    a_1(r,l)=\binom{L-r-1}{M-r+l}, \ a_{k+1}(r,l)=a_{k}(r,l+1)-a_{k}(r,l)\ (k,r=1,2,\ldots,\ l=0,1,\ldots).
    \label{eq:recurrence_relation}
\end{equation}
Then, the last line of \eqref{eq:singlet_approx} is written as $|a_r (r,0)|$. We consider the thermodynamic limit $L,M\to \infty$ such that $k$, $r$, $l$, and the filling $\nu=M/L$ are kept constant. In this limit, we see below that $a_k (r,l)$ is expressed as follows.
\begin{equation}
    a_k (r,l)=\left[\nu (1-\nu) (1-2\nu)^{k-1}+O(1/L) \right]\binom{L-r+k}{M+l-r+k}
    \label{eq:induntion}
\end{equation}
We prove Eq. \eqref{eq:induntion} by induction.

(i) When $k=1$,
\begin{equation}
        a_1(r,l)
        = \binom{L-r-1}{M-r+l} =\frac{\binom{L-r-1}{M-r+l}}{\binom{L-r+1}{M+l-r+1}}\binom{L-r+1}{M+l-r+1}
        =\frac{(M-r+l)(L-M-l)}{(L-r+1)(L-r)}\binom{L-r+1}{M+l-r+1}.
\end{equation}
Since $M/L=\nu$, $l/L=O(1/L)$, and $r/L=O(1/L)$, we have
\begin{equation}
a_1(r,l)= [\nu (1-\nu) +O(1/L)]\binom{L-r+1}{M+l-r+1}.
\end{equation}
So we obtain Eq. \eqref{eq:induntion} when $k=1$.

(ii) We assume \eqref{eq:induntion} when $k=k^\prime$. Then, from \eqref{eq:recurrence_relation},
\begin{equation}
\begin{split}
a_{k^\prime+1} (r,l)
&=a_{k^\prime} (r,l+1)-a_{k^\prime} (r,l)\\
&=\left[\nu (1-\nu) (1-2\nu)^{k^\prime-1}+O(1/L) \right]\left[\binom{L-r+k^\prime}{M+l-r+k^\prime+1}-\binom{L-r+k^\prime}{M+l-r+k^\prime}\right] \\
&=\left[\nu (1-\nu) (1-2\nu)^{k^\prime-1}+O(1/L) \right]\frac{\left[\binom{L-r+k^\prime}{M+l-r+k^\prime+1}-\binom{L-r+k^\prime}{M+l-r+k^\prime}\right]}{\binom{L-r+k^\prime+1}{M+l-r+k^\prime+1}}\binom{L-r+k^\prime+1}{M+l-r+k^\prime+1}\\
&=\left[\nu (1-\nu) (1-2\nu)^{k^\prime-1}+O(1/L) \right]
\left[
\frac{(L-M-l)-(M+l-r+k^\prime+1)}{L-r+k^\prime+1}
\right]
\binom{L-r+k^\prime+1}{M+l-r+k^\prime+1}.
\end{split}
\end{equation}
When $k'/L,l/L, r/L=O(1/L)$, we have
\begin{equation}
\begin{split}
a_{k^\prime+1} (r,l)
&=\left[\nu (1-\nu) (1-2\nu)^{k^\prime-1}+O(1/L) \right]
\left[
(1-2\nu) +O(1/L)
\right]
\binom{L-r+k^\prime+1}{M+l-r+k^\prime+1} \\
&=\left[\nu (1-\nu) (1-2\nu)^{k^\prime}+O(1/L) \right]
\binom{L-r+k^\prime+1}{M+l-r+k^\prime+1}.
\end{split}
\end{equation}
Thus we have Eq. \eqref{eq:induntion} when $k=k^\prime +1$.

From (i) and (ii), we have \eqref{eq:induntion} for all $k$ $(\ll L)$. Therefore, when $r\ll L$, we obtain
\begin{equation}
    |\langle \hat{\eta}^\dagger_x \hat{\eta}_y \rangle^L_M|=\left|\frac{a_r(r,0)}{\binom{L}{M}}\right|= \nu (1-\nu)|1-2\nu|^{r-1} +O(1/L).
\end{equation}
\end{widetext}

\section{CALCULATION OF THE SINGLET CORRELATION FUNCTION USING A SPIN CHAIN}
\label{sec:singlet_spin}
\begin{figure}[H]
  \centering
  \includegraphics[width=1\linewidth]{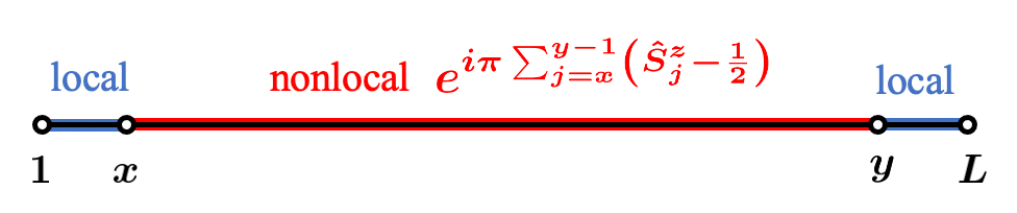}
  \caption{The singlet correlation function in the language of spin operators is defined by a nonlocal operator when $N$ is odd. However, when $\hat{S}^z_\mathrm{tot}$ is fixed and $r=|x-y| \simeq 0 \text{ or } L$, it can be calculated by local operators defined around the two edges.}
  \label{fig:local_nonlocal}
\end{figure}

We map $\ket{\Phi^L_M}$ to a state on a spin chain using \eqref{eq:spin_map}. First, $\ket{\Phi^{L}(\alpha)}$ is mapped to
\begin{equation}
  \ket{\Phi^{L}(\alpha)_\mathrm{spin}}:=\bigotimes_{j=1}^{L}\left(|\downarrow\rangle_{j}+\alpha|\uparrow\rangle_{j}\right).
  \label{eq:spin_chain}
\end{equation}
Thus $\ket{\Phi^L_M}$ is mapped to a ferromagnetic state $\ket{L/2,M/2}$ with $\hat{\bm{S}}^2_\mathrm{tot}\ket{L/2,M/2}=L/2(L/2+1)\ket{L/2,M/2}$ and $\hat{S}^z_\mathrm{tot}\ket{L/2,M/2}=M/2\ket{L/2,M/2}$. The operator $P \hat{\eta}^\dagger_x \hat{\eta}_y P$ is mapped to
\begin{equation}
  \begin{cases}
    (-1)^{x+y} \hat{S}_x^+ \hat{S}_y^- & \text{ when } N \text{ is even, } \\
    \hat{S}_x^+ e^{i \pi \sum_{j=x}^{y-1}\left(\hat{S}_j^z-\frac{1}{2}\right)} \hat{S}_y^- & \text{ when } N \text{ is odd. }
  \end{cases}
  \label{eq:spin_chain_correlation}
\end{equation}
When $N$ is even, the expectation value of $\hat{S}_x^+ \hat{S}_y^-$ does not depend on $x$ and $y$. When $N$ is odd, due to the nonlocal term $e^{i \pi \sum_{j=x}^{y-1}\left(\hat{S}_j^z-\frac{1}{2}\right)}$, the singlet correlation function decays when $r=|x-y|$ increases. However, if $\hat{S}^z_\mathrm{tot}$ is fixed to $M/2$, we have
\begin{equation}
\begin{split}
  &e^{i \pi \sum_{j=x}^{y-1}\left(\hat{S}_j^z-\frac{1}{2}\right)} \ket{L/2,M/2} \\
  &= e^{i \pi \left[\left(\frac{M}{2}-\frac{L}{2}\right)-\sum_{j=y}^{L}\left(\hat{S}_j^z-\frac{1}{2}\right)- \sum_{j=1}^{x-1}\left(\hat{S}_j^z-\frac{1}{2}\right)\right]\ket{L/2,M/2}}.
\end{split}
\end{equation}
Therefore, if $r=|x-y| \simeq L$, this operator is written as a product of spin operators defined on a few sites around the edges, as shown in Fig.~\ref{fig:local_nonlocal}. In particular, when $x=1$ and $y=L$,
\begin{equation}
\begin{split}
 &\hat{S}_{1}^{+} e^{i \pi \sum_{j=1}^{L-1}\left(\hat{S}_j^z-\frac{1}{2}\right)} \hat{S}_{L}^{-}\ket{L/2,M/2} \\
 &=-e^{i \pi\left(\frac{M-L}{2}\right)} \hat{S}_{1}^{+} \hat{S}_{L}^{-}\ket{L/2,M/2}.
\end{split}
\end{equation}
Therefore, the correlation between two edges $1, L$ does not decay at any filling $0<\nu=M/L<1$. This is an interpretation of the revival of the singlet correlation function when $r \simeq L$.

\section{DERIVATION OF EQS. \eqref{eq:norm_Phi_alpha} and \eqref{eq:singlet_correlation_Phi_alpha}}
\label{sec:derivation_19}
Here we derive Eqs. \eqref{eq:norm_Phi_alpha} and \eqref{eq:singlet_correlation_Phi_alpha}. From \eqref{eq:spin_chain},
\begin{equation}
  \braket{\Phi^{L}(\alpha)}=\braket{\Phi^{L}(\alpha)_\mathrm{spin}}=(1+\alpha^2)^L.
\end{equation}

Using \eqref{eq:spin_chain}, \eqref{eq:spin_chain_correlation},
${\hat{S}_j^+ \left(\ket{\downarrow}_{j}+\alpha\ket{\uparrow}_{j}\right)=\ket{\uparrow}_{j}}$,
${\hat{S}_j^- \left(\ket{\downarrow}_{j}+\alpha \ket{\uparrow}_{j}\right)=\alpha\ket{\downarrow}_{j}}$,
and
${e^{i \pi \left(\hat{S}_j^z-\frac{1}{2}\right)}\left(\ket{\downarrow}_{j}+\alpha\ket{\uparrow}_{j}\right)={-\ket{\downarrow}_{j}+\alpha\ket{\uparrow}_{j}}}$,
we have
\begin{equation}
\begin{split}
      &\left\langle\Phi^{L}(\alpha)\left|\hat{\eta}_{x}^{\dagger} \hat{\eta}_{y}\right| \Phi^{L}(\alpha)\right\rangle \\
      &=\bra{\Phi^{L}(\alpha)_\mathrm{spin}} (-1)^{x+y} \hat{S}_x^+ \hat{S}_y^- \ket{\Phi^{L}(\alpha)_\mathrm{spin}} \\
      &= (-1)^{x+y} \alpha^2 (1+\alpha^2)^{L-2}
\end{split}
\end{equation}
for $N$ even and
\begin{equation}
\begin{split}
  &\left\langle\Phi^{L}(\alpha)\left|\hat{\eta}_{x}^{\dagger} \hat{\eta}_{y}\right| \Phi^{L}(\alpha)\right\rangle \\
  &=\bra{\Phi^{L}(\alpha)_\mathrm{spin}} \hat{S}_x^+ e^{i \pi \sum_{j=x}^{y-1}\left(\hat{S}_j^z-\frac{1}{2}\right)} \hat{S}_y^- \ket{\Phi^{L}(\alpha)_\mathrm{spin}} \\
  &=(-1)^{x+y} \alpha^{2}\left(1+\alpha^{2}\right)^{L-r-1}\left(1-\alpha^{2}\right)^{r-1}
  \end{split}
\end{equation}
for $N$ odd.

\section{SU(3) SYMMETRIC HAMILTONIAN WHERE $\eta$-PAIRING AND $\eta$-CLUSTERING EIGENSTATES COEXIST}
\label{sec:coexist}

When $N=3$, three-body $\eta$-clustering states are eigenstates of $\hat{H}_\mathrm{OBC}$, but two-body $\eta$-pairing states are not. 
Here we construct a model in which these two kinds of states are exact energy eigenstates.
We consider a chain of three-component fermions with $L$ lattice sites, and assume that $L$ is even. Let us consider the following Hamiltonian
with the periodic boundary conditions:
\begin{equation}
\begin{split}
\hat{H}^{(3)}=&-t \sum_{x=1}^{L} \left( \hat{T}_{x,x+1}+\frac{9}{2}\hat{V}_{x,x+1}\right)\\
&+U\sum_{x=1}^{L} \hat{n}_x\left(\hat{n}_x-3\right),
\end{split}
\end{equation}
where $t \in \mathbb{R}$ is the hopping amplitude and $U \in \mathbb{R}$ is the strength of interaction.
As we have seen in Remark 1, the $\eta$-clustering states $\ket{\Phi^{L}_{M}}$ are exact eigenstates of the Hamiltonian $-t \sum_{x=1}^{L}  \hat{T}_{x,x+1}
+U\sum_{x=1}^{L} \hat{n}_x\left(\hat{n}_x-3\right)$ when $M$ is odd. The states $\ket{\Phi^{L}_{M}}$ are also the ground states of the Hamiltonian $\sum_{x=1}^{L}\hat{V}_{x,x+1}$. We have seen this in Sec. \ref{sec:parent} in the case of open boundary conditions, and the extension
to the periodic case is straightforward. Therefore, the $\eta$-clustering states $\ket{\Phi^{L}_{M}}$ are exact eigenstates of the Hamiltonian $\hat{H}^{(3)}$ when $M$ is odd.
Next, we see that two-body $\eta$-pairing states are also eigenstates of the Hamiltonian. First, we define two-body $\eta$-operators as
\begin{align}
    (\hat{\eta}^{\sigma,\tau})^{\dagger} &=\sum_{x=1}^L e^{i\pi x} \hat{c}^\dagger_{x,\sigma}\hat{c}^\dagger_{x,\tau}.
\end{align}
Note that $(\hat{\eta}^{\tau,\sigma})^{\dagger}=-(\hat{\eta}^{\sigma,\tau})^{\dagger}$ and hence $(\hat{\eta}^{\sigma,\sigma})^{\dagger}=0$. Then, we define generalized $\eta$-pairing states \footnote{These states are considered in Ref. \cite{nakagawa_eta_unpublished}, but the Hamiltonian is different from $H^{(3)}$.}~as
\begin{equation}
\begin{split}
&\ket{\psi(M_{1,2},M_{2,3},M_{3,1})}\\
&=\left\{\left(\hat{\eta}^{1,2}\right)^\dagger\right\}^{M_{1,2}} \left\{\left(\hat{\eta}^{2,3}\right)^\dagger\right\}^{M_{2,3}} \left\{\left(\hat{\eta}^{3,1}\right)^\dagger\right\}^{M_{3,1}} \ket{0}.
    \label{eq:eta_two_su3}
    \end{split}
\end{equation}
Here, $M_{1,2}$, $M_{2,3}$, and $M_{3,1}$ are non-negative integers that satisfy $0\leq M_{1,2}+M_{2,3}+M_{3,1}\leq L$. In Ref.~\cite{nakagawa_eta_unpublished}, it is proven that $\ket{\psi(M_{1,2},M_{2,3},M_{3,1})}$ are eigenstates of the SU(3) Hubbard model. To see that $\ket{\psi(M_{1,2},M_{2,3},M_{3,1})}$ are eigenstates of $H^{(3)}$, we first consider the case where $M_{2,3}=M_{3,1}=0$. The Hamiltonian $H^{(3)}$, when restricted to the subspace where there are no fermions with $\sigma=3$, behaves as if it were
\begin{align}
&\hat{H}^{(3)}|_{\sigma=1,2} \\
&= -t\sum_{x=1}^{L} \sum_{\sigma=1}^{2}(\hat{c}^\dagger_{x,\sigma} \hat{c}_{x+1,\sigma}+\text{h.c.})\label{eq:h3_1v}\\
&+U\sum_{x=1}^{L} \hat{n}_x\left(\hat{n}_x-2\right)\label{eq:h3_2v}\\
& -2t\sum_{x=1}^{L} \left\{\frac{1}{2}\hat{\overline{c}}^\dagger_{x,3} \hat{\overline{c}}_{x+1,3}+\text{h.c.})\right.\nonumber\\
&\ \ \ \ \ \ \ \ \ \ \ \ \ \  -\left. \frac{1}{4} \hat{n}_x \hat{n}_{x+1}+\frac{1}{4} (\hat{n}_x+\hat{n}_{x+1}) \right\}\label{eq:h3_3v}\\
&+\left(\frac{t}{2}-U\right)\sum_{x=1}^{L} \hat{n}_x.\label{eq:h3_4v}
\end{align}
Here we used the fact that the three-body hopping term in $V_{x,x+1}$ vanishes in this subspace. In the subspace, $\ket{\psi(M_{1,2},0,0)}$ can be seen as ordinary $\eta$-pairing states. Thus $\ket{\psi(M_{1,2},0,0)}$ is an eigenstate of \eqref{eq:h3_1v} and \eqref{eq:h3_2v}, which is the SU(2) Hubbard model~\cite{yang__1989}. The state is an eigenstate of the term \eqref{eq:h3_3v}, because this term is proportional to \eqref{eq:hv} with $N=2$ if we replace $\hat{\overline{c}}^\dagger_{x,3}\ (\hat{\overline{c}}_{x,3})$ with $\hat{\eta}^\dagger_x\ (\hat{\eta}_x)$. Finally, we see that $\ket{\psi(M_{1,2},0,0)}$ is an eigenstate of \eqref{eq:h3_4v}, because \eqref{eq:h3_4v} is constant if the number of particle is fixed. Therefore, $\ket{\psi(M_{1,2},0,0)}$ is an eigenstate of $H^{(3)}$ for all $M_{1,2}$.

We now move on to the case where $M_{2,3}$ or $M_{3,1}$ is nonzero. To this end, we introduce the operators $\hat{F}^{\sigma, \tau}= \sum_{x=1}^{L}\hat{c}^\dagger_{x,\sigma}\hat{c}_{x,\tau}$. Here, $\hat{F}^{\sigma, \sigma}$ is the total number operator of fermions with flavor $\sigma$, while $\hat{F}^{\sigma, \tau}$ ($\sigma \ne \tau$) are flavor-raising and lowering operators.
Since $\hat{F}^{\sigma, \tau}$ operators commute with $H^{(3)}$, if a state $\ket{\phi}$ is an eigenstate of $H^{(3)}$, $\hat{F}^{\sigma, \tau}\ket{\phi}$ is also an eigenstate of $H^{(3)}$.
By using the commutation relations
\begin{align}
   \left[(\hat{\eta}^{\sigma,\tau})^\dagger,\ \hat{F}^{\mu, \nu}\right]&=\delta_{\sigma,\nu}(\hat{\eta}^{\tau,\mu})^\dagger- \delta_{\tau,\nu}(\hat{\eta}^{\sigma,\mu})^\dagger, \\
   \left[(\hat{\eta}^{\sigma,\tau})^\dagger,\ (\hat{\eta}^{\mu, \nu})^\dagger\right]&=0,
\end{align}
one finds
\begin{equation}
\begin{split}
   &\hat{F}^{3, 1} \ket{\psi(M_{1,2},M_{2,3},M_{3,1})}\\
   =&-M_{1,2}\ket{\psi(M_{1,2}-1,M_{2,3}+1,M_{3,1})}
   \label{eq:ind1}
\end{split}
\end{equation}
and
\begin{equation}
\begin{split}
   &\hat{F}^{3, 2} \ket{\psi(M_{1,2},M_{2,3},M_{3,1})}\\
   =&-M_{1,2}\ket{\psi(M_{1,2}-1,M_{2,3},M_{3,1}+1)}.
    \label{eq:ind2}
\end{split}
\end{equation}
Thus we see that
\begin{equation}
\begin{split}
   &\ket{\psi(M_{1,2},M_{2,3},M_{3,1})}\\
   &= c(M_{1,2},M_{2,3},M_{3,1}) (\hat{F}^{3, 2})^{M_{3,1}}(\hat{F}^{3, 1})^{M_{2,3}}\\
   &\times\ket{\psi(M_{1,2}+M_{2,3}+M_{3,1},0,0)},
    \label{eq:ind3}
\end{split}
\end{equation}
where
\begin{equation}
    c(M_{1,2},M_{2,3},M_{3,1})=\frac{(-1)^{M_{2,3}+M_{3,1}}M_{1,2}!}{(M_{1,2}+M_{2,3}+M_{3,1})!}.
\end{equation}
Since $\ket{\psi(M_{1,2}+M_{2,3}+M_{3,1},0,0)}$ is an eigenstate of
$H^{(3)}$ as shown before, $\ket{\psi(M_{1,2},M_{2,3},M_{3,1})}$ is an eigenstate of $H^{(3)}$ for all $M_{1,2}$, $M_{2,3}$ and $M_{3,1}$.

\bibliographystyle{apsrev4-1}
\bibliography{thesis}
\end{document}